\documentclass[preprint,showpacs,preprintnumbers,amsmath,amssymb]{revtex4}

\usepackage{graphicx}
\usepackage{dcolumn}
\usepackage{bm}

\begin{document}

\preprint{APS/}

\title{Characterizing heralded single photons with an all-fiber source of photon pairs}

\author{Lei Yang, Xiaoxin Ma, Xueshi Guo, Liang Cui and Xiaoying Li}%
\email[]{xiaoyingli@tju.edu.cn}

\address {College of Precision Instrument and
Opto-electronics Engineering, Tianjin University, \\Key Laboratory of Optoelectronics Information Science and Technology, Ministry of Education, Tianjin, 300072,
P. R. China}

\begin{abstract}
Based on the signal and idler photon pairs produced in a piece of high nonlinear fiber by a pulsed pump, we characterize the heralded single photon source from both the theoretical and experimental aspects. In the theory model, started from the derived expression of Bogoliubov transformation for a broadband pulsed pump, the second-order intensity correlation function $g_c^{(2)}(0)$, heralding efficiency $H$, indistinguishability and brightness of the heralded single photons as a function of source parameters are analyzed and discussed. In the experiments, using several kinds of combinations of the source parameters, the values of $g_c^{(2)}(0)$ and $H$ are measured and compared. The experimental results are consistent with the theoretical predictions. The investigations are useful for optimizing the parameters and for developing a single photon source suitable for quantum information processing.
\end{abstract}

\maketitle

\section{Introduction}

Generation of single photons is not only important for the fundamental test of quantum mechanics, but also crucial for quantum metrology and quantum information processing~\cite{lounis05}. An ideal single photon source would produce completely characterized single photons on demand. However, all the available sources fall short of this ideal. To evaluate the quality of a single photon source, the following three criteria are often taken into account: (i) photon statistics, represented by the second-order intensity correlation function; (ii) efficiency, characterized by the probability of delivering a photon in response to a request; and (iii) indistinguishability, described by its mode structure. In practice, the requirements depend on its specific applications. For example, for linear optics quantum computing involving interference between photons from multiple sources~\cite{Knill01}, the indistinguishability is important; while for quantum key distribution (QKD), which works well with a few modes, the main concerns are the photon statistics and efficiency~\cite{Gisin02}.

In general, single photon sources can be classified into two categories: deterministic source and heralded source.
The first type relies on the fact that a single isolated quantum system can emit only one photon at each time it is excited\cite{kuhm02,brunel99,kurtsiefer00}. The second uses the spontaneous parametric emission of distinguishable signal and idler photon pairs, the detection of the one member of the pair indicates the existence of its twin photon. Since the physical process of parametric emission is constrained by the energy and momentum conservations, the detected location of the heralding photon defines the location of its twin. This is a significant advantage over the other type of source.

The heralded single photon sources (HSPSs), which are inherently
probabilistic, have traditionally been based on spontaneous parametric down conversion (SPDC) in $\chi ^{(2)}$ crystals. Since the first conditional preparation of single photon state reported by Hong and Mandel~\cite{hong86}, this kind of source has been characterized from various aspects, and significant progress has been achieved in improving collection efficiency, in optimizing the photon statistics, and in engineering the mode structure~\cite{Castelletto08}. Recently, experiments and theoretical investigations have shown that photon pairs generated in optical fiber by $\chi^{(3)}$-based spontaneous four wave mixing (SFWM) can also be used to develop the  HSPS\cite{Fan08,cohen09,mcmillan09,Garay-Palmett-OE2007}. Moreover, the fiber based HSPS has the advantages in single spatial mode, low collection loss, compact physical
size, low pump power and compatibility with an optical
network.

The spontaneous parametric emissions can be realized by using either a continuous wave laser or a mode-locked pulsed laser. For the quantum information
processing applications requiring synchronization~\cite{Knill01,Banaszek04}, photon pais with creation time defined by the pulsed pump are often preferred. The HSPS based on the pulsed SPDC has been experimentally demonstrated for some time, but its comprehensive theoretical description is not available until recently~\cite{Ou97,Grice97,Vicent10,Branczyk10}. In Ref.~\cite{Vicent10}, the brightness and purity of HSPS are analyzed by taking both the spectral and spatial degree of freedoms into account; while in Ref.~\cite{Branczyk10}, the dependence of the three criteria for evaluating the quality of HSPS are simulated by expanding the output state of the SPDC in $\chi ^{(2)}$ crystal based waveguide to second order in photon number. In contrast, for the the HSPS based on the SFWM in fibers, which is usually pumped with a pulsed laser, there has been a few experimental demonstrations~\cite{Fan08,cohen09,mcmillan09}, however, the detailed characterizations have not been done yet.

In this paper, using the signal and idler photon pairs generated in 20-meter long high nonlinear fiber (HNLF) by a pulsed pump in 1550 nm band via SFWM, we carefully characterize the heralded single photon source from the theoretical and experimental aspects. In the theory model, to take the multi-photon emission into account in a simpler way, the evolution of signal and idler fields is calculated in Heisenberg picture. Based on the derived expression of the Bogoliubov transformation for a broadband pulsed pump, the fomular of the second-order intensity correlation function $g_c^{(2)}(0)$ and heralding efficiency $H$ of the HSPS are obtained with analytic method for the first time. In the experiments, exploiting the advantaged conditions provided by the various kinds of commercially available off-the-shelf fiber components, the values of $g_c^{(2)}(0)$ and $H$ are measured and analyzed by using an all fiber HSPS, and the results agree with theoretical predictions.

The rest of the paper is organized as follows. Section 2 gives the conceptual representation of the HNLF-based HSPS. Section 3 works out a theoretical frame suitable for describing the HSPS, the dependence of the photon statistics, heralding efficiency, indistinguishability and brightness upon the source parameters are theoretically analyzed and discussed. Section 4 demonstrates the experiments of the all fiber HSPS with several kinds of combination of source parameters, and the experimental results are compared with the theoretically calculated results with no fitting parameters. Finally, we give a brief conclusion and discuss the possible improvements.

\section{Conceptual representation of the heralded single photon source}

The conceptual representation of the scheme for generating heralded single photons is shown in Fig. 1. The HSPS is based on the photon pairs generated in HNLF via SFWM by a pulsed pump with a pulse width and
repetition rate of about 10 picoseconds and 41 MHz, respectively. When the central wavelength of the pump pulses is in the anomalous-dispersion regime,
the parametric process is phase-matched and the probability of SFWM is significantly
enhanced. In this process, two
pump photons at frequency $\omega _{p1}$ and $\omega _{p2}$, respectively, scatter through the Kerr ($\chi ^{(3)}$)
nonlinearity of the fiber to create
energy-time entangled signal and idler photons at frequencies $\omega _s$ and $\omega _i$, respectively, such that $\omega _{p1}+\omega _{p2}=\omega _s+\omega
_i$. The photons created in pairs are originated from the vacuum fields at frequencies $\omega _s$ and $%
\omega _i$, $a_{in}(\omega_s)$ and $a_{in}(\omega_i)$, respectively.  At the output port of the HNLF, photon pairs  predominantly
co-polarized with the pump photons are described by field operators $a_{out}(\omega_s)$ and $a_{out}(\omega_i)$.

To measure the signal and idler photons, one must effectively suppress the pump photons from
reaching the detector, so the output of HNLF is passed through a dual band filter F2. Signal and idler photons, described by the field operators, $b(\omega_s)$ and $b(\omega_i)$, are then detected by the photon counting system consisting of three InGaAs/InP avalanche photodiode-based single photon detectors (SPDs). The SPDs are operated in the gated Geiger-mode, the gate pulses arrive at a rate of about $2.58$\,MHz, which is $1/16$ of the repetition rate of the pump pulses, and the dead time of the gate
is set to be 10 $\mu$s. The pulse widths of the gate for the SPDs are 2.5 ns. The timing of the gate pulses are adjusted by a digital delay generator to coincide with the arrival of signal and idler photons.

The photons in idler band are detected by SPD1, whose detection signals are used to herald the existence of single photons in signal band. To characterize the HSPS, photons in signal band propagate through a 50/50 fiber coupler and then detected by SPD2 and SPD3, respectively. To measure the second order coherence function and heralding efficiency, besides recording the single counts of each SPD, the two-fold and three-fold coincidence between SPD1, SPD2, and SPD3 also need to be recorded.

It is worth noting that in our experiment, the detuning between signal (idler) and pump photons is about 0.8 THz. In this case, the photons in signal and idler bands originated from self phase modulation are negligible. However, the SFWM in HNLF is inevitably accompanied by Raman scattering ~\cite{Li04}. For simplicity, only the SFWM process is considered in our theoretical analysis.

\section{Theory}

To characterize the most important feature of a HSPS --- the second order intensity correlation function, the theoretical model should take the multi-photon emission of the source into account. This issue has been investigated in Ref.\cite{razavi09}, in which the HSPS relies on continuous-wave SPDC in $\chi ^{(2)}$ crystal, and the analysis of second-order intensity correlation properties is based on the signal and idler fields evolved via the Bogoliubov transformation in single mode form~\cite{Shapiro94}. To work out a theoretical frame suitable for describing our fiber based HSPS, an expression of the Bogoliubov transformation for broadband pulsed pump is derived, and the theory in Ref.\cite{razavi09} is expanded to a multi-mode form.

The interaction Hamiltonian in the volume within the fiber that is capable of  describing the interaction of SFWM is~\cite{Chen05,Alibart06}
\begin{eqnarray}
\label{1}
H_{I}=\alpha \chi ^{(3)}\int {dV}\left( {%
E_{p1}^{+}E_{p2}^{+}E_{s}^{-}E_{i}^{-}}+h.c.\right),
\label{}
\end{eqnarray}
where $\alpha $ is a constant determined by experimental details, ${E_{pi}^{+}}$ ($i=1,2$) is the positive frequency electric-field operator of pumps, ${%
E_{s}^{-}}$ and ${%
E_{i}^{-}}$ are the negative frequency electric-field operators of signal and idler fields, respectively,
 and $h.c.$ stands for Hermitian conjugate. The integral is taken over the entire volume of
interaction $V$. Since we assume all the optical fields are linearly co-polarized, the tensorial nature of the nonlinear electric susceptibility $\chi ^{(3)}$ can be ignored.

In the model, the strong pump pulse propagating along the fiber (denoted as z direction) with a Gaussian shaped spectrum remains classical, it can be written as
\begin{eqnarray}
\label{2}
E_{pi}^ + = E_{p0} e^{ - i\gamma P_p z} \int {d\omega _{pi} e^{ - (\omega _{pi} -
\omega _{p0} )^2 /2\sigma _p^2 } e^{ik_{pi} z - i\omega _{pi} t}},
\end{eqnarray}
where
$P_{p}=2\sqrt{\pi }A_{eff}\varepsilon _{0}cn(\omega)\sigma _{p}^{2}E_{p0}^{2}$
is the peak power of the pump pulse, which is treated as a constant under the undepleted pump approximation,  $A_{eff}$ and $\varepsilon _{0}$ denote the effective mode area of
the optical fiber and the vacuum permittivity, respectively; $c$ and $n(\omega)$ are the speed of light in vacuum and refractive index of the fiber, respectively; $\sigma
_{p}$ and $\omega _{p0}$ are the optical bandwidth and the central frequency
of the pump pulses, $k_{pi}$ is the wave vector of pump, and $\gamma =3\omega_{p0} Re(\chi ^{(3)})/(4n(\omega)^2c^2\varepsilon _0 A_{eff})$ is the nonlinear coefficient. The signal and idler fields are quantized electromagnetic fields
as given by the following
multi-mode expansion:
\begin{eqnarray}
\label{3}
E_{j}^{-}=\int {d\omega _{j}}\sqrt{\frac{\hbar \omega _{j}}{2\varepsilon
_{0}V_{Q}}}\frac{{a^{\dag }(\omega _{j})}}{n(\omega)}{e^{-ik_{j}z+i\omega _{j}t}},%
~~~(j=s,i),
\label{}
\end{eqnarray}
where $a^\dag (\omega _j )$ is the creation operator of the field at frequency $\omega _j$, $k_{j}$ is its wave vector, and $V_{Q}$ defines the quantization volume.

Substituting Eqs. (\ref{2})-(\ref{3}) into Eq. (\ref{1}), and carrying out the integration over the volume, the interaction Hamiltonian can be written as:
\begin{eqnarray}
\label{4}
H_I=\frac{\alpha ^{\prime }{\gamma P_pL}}{{\sigma _p^2}}\int {d\omega
_{p1}d\omega _{p2}d\omega _sd\omega _ia^{\dag }(\omega _s)a^{\dag
}(\omega _i)\mathrm{sinc}\left( \ {\frac{{\Delta kL}}2}\right) }e^{\frac{{i \Delta kL}}2} \cr \times e^{-\left\{ {(\omega _{p1}-\omega _{p0})^2+(\omega _{p2}-\omega
_{p0})^2}\right\} /2\sigma _p^2}e^{-i(\omega _{p1} +\omega _{p2}-\omega
_s-\omega _i)t} +h.c.,
\end{eqnarray}
where  $\alpha ^{\prime }=\alpha A_{eff}\hbar c/\left( 3\sqrt{\pi }\varepsilon
_{0}nV_Q\right) $, $L$ is the
length of the optical fiber, and $\Delta k=k_{p1}+k_{p2}-k_s-k_i-2\gamma P_p$ is the phase-mismatching term. For the pulsed pump with the central wavelength in the anomalous-dispersion regime of HNLF, the bandwidth of the $\mathrm{sinc}$
function is usually greater than 10 nm, which is 10 times wider than that of filter F2 used in the experiments. Therefore, we can assume
$\mathrm{sinc}\left( {\frac{{\Delta kL}}{2}}\right) \approx 1$.

To obtain the unitary evolution operator $U = \exp \left\{ {\frac{1} {{i\hbar }}\int {dtH_I } } \right\} $, we need to carry out the integration over time and over all the possible combinations of $\omega _{p1}$ and $\omega _{p2}$ within the pump bandwidth. Since the integration over time gives rise to the $\delta $ function in $\Delta \omega =\omega _{p1}+\omega _{p2}-\omega _s-\omega _i$ to guarantee the energy conservation at single photon level, the evolution operator can be expressed as:
\begin{eqnarray}
\label{5}
U= \exp \left\{ {%
\frac{G}{
\sigma _{p}} \int {d\omega _sd\omega _i\phi (\omega _s,\omega _i)a_{}^{\dag }(\omega
_s)a_{}^{\dag }(\omega _i)+h.c.}}\right\}
\end{eqnarray}
with
\begin{eqnarray}
\label{6}
\phi (\omega _s ,\omega _i ) = \exp \{ - \frac{1}{{4\sigma _p^2 }}(\omega _s
+ \omega _i - 2\omega _{p0} )^2 \},
\end{eqnarray}
where $G =-i2\pi \sqrt{\pi }\alpha ^{\prime }\gamma P_pL/\hbar $, is proportional to the gain of SFWM.

In the Heisenberg picture, the field operator of signal (idler) beam at the output of HNLF can be written as
\begin{eqnarray}
\label{bogo}
a_{out}(\omega _{s(i)} )=U^{\dag }a_{in}(\omega _{s(i)})U = \int {d\omega^{\prime}_{s(i)} h_1 (\omega^{\prime}_{s(i)} ,\omega _{s(i)} )a_{in} (\omega^{\prime}_{s(i)} )%
} + \int {d\omega_{i(s)} h_2 (\omega_{i(s)},\omega _{s(i)} )a_{in}^\dag (\omega_{i(s)} )}
\end{eqnarray}
with
\begin{eqnarray}
\label{8}
\left| \int d \omega^{\prime}_{s(i)} h_1(\omega^{\prime}_{s(i)} ,\omega _{s(i)})\right| ^2-\left|\int d \omega_{i(s)} h_2(\omega_{i(s)} ,\omega _{s(i)})\right| ^2=1,
\end{eqnarray}
where the complex functions $h_1 (\omega^{\prime}_{s(i)} ,\omega
_{s(i)} )$ and $h_2 (\omega_{i(s)} ,\omega _{s(i)} )$ are given by
\begin{eqnarray}
\label{9}
h_1 (\omega^{\prime}_{s(i)} ,\omega _{s(i)} ) =
\sum\limits_{n = 0}^\infty {\frac{{1 }}{{%
\sqrt {2n}{(2n)!} }}\frac{{\left| G \right|^{2n} }}{2\sqrt{\pi}\sigma _p }\exp \{ - \frac{1}{{%
4\sigma _p^2 \times 2n}}(\omega^{\prime}_{s(i)} - \omega _{s(i)} )^2 \} }
\end{eqnarray}
and
\begin{eqnarray}
\label{10}
h_2 (\omega_{i(s)} ,\omega _{s(i)} ) = \sum\limits_{n = 0}^\infty {\frac{{1 }}{{\sqrt {2n + 1}{(2n + 1)!} }}\frac{{%
\left| G \right|^{2n } {G}}}{2\sqrt{\pi}\sigma _p }\exp \{ - \frac{1}{{4\sigma _p^2
\times (2n + 1)}}(\omega_{i(s)} + \omega _{s(i)} - 2\omega _{p0} ^2 \} }.
\end{eqnarray}
The field operators $a_{out}(\omega _{s} )$ and $a_{out}(\omega _{i} )$ commute with each other, and individually obey the commutation relation
$[a_{out}(\omega ),a_{out}^\dag (\omega ^{\prime })] = \delta (\omega - \omega  ^{\prime })$. Therefore, the transformation represented by Eqs.(\ref{bogo})-(\ref{10}) is equivalent to the Bogoliubov transformation. To our knowledge this is the first formulation of such a general expression for broadband pulsed pump.

In the low gain regime, which is of interest to us, $a_{out}(\omega _{s(i)})$ can be expressed in the following form:
\begin{eqnarray}
\label{11}
a_{out}(\omega _{s(i)})={a_{in}(\omega _{s(i)})}+\frac{G}{
\sigma _{p}} \int {d\omega_{i(s)} \exp \{-\frac{1}{{%
4\sigma _{p}^{2}}}(\omega_{i(s)} +\omega _{s(i)}-2\omega _{p0})^{2}\}a_{in}^{\dag
}(\omega_{i(s)} )}+o(G).
\end{eqnarray}
After passing through the dual band filter F2, which is described by the function
$f_{j}(\omega _{j})=\exp \{-\frac{{(\omega _{j}-\omega _{j0})^{2}}}{{2\sigma
_{j}^{2}}}\}$ ~(j=s,i),
where $\omega _{j0}$ and $\sigma
_{j}$ respectively denote its central frequency and bandwidth, the field operators of signal and idler photons involve into
\begin{eqnarray}
\label{12}
b(\omega _j)=\sqrt{\eta _j}{f}_{{j}}{(\omega }_{{j}}{)a(\omega _j)},
\end{eqnarray}
where  $\eta _j$ (j=s,i) is transmission efficiency of F2.

Using the field operators described by Eqs (\ref{11}) and (\ref{12}), it is straightforward to obtain the electric-field operators at the SPDs, with which the photon statistics and heralding efficiency of the HSPS can be calculated. If the $k$th pump pulse is centered at $t_{0k}$ in time domain, the idler field detected by SPD1 (per pulse) can be written as
\begin{eqnarray}
\label{13}
E_1^{+}(t-t_{0k})=\frac {\sqrt{\eta _1}}{\sqrt{2\pi }}\int {d\omega _ib(\omega _i)e^{-i\omega _i(t-t_{0k})}
},
\end{eqnarray}
where $\eta _1$ is quantum efficiency of SPD1; and the signal fields, passed through the 50/50 beam splitter and detected by SPD2 (SPD3), can be written as
\begin{eqnarray}
\label{14}
E_{2(3)}^{+}(t-t_{0k})=\frac {\sqrt{\eta _{2(3)}}}{\sqrt{4\pi }}\int {d\omega
_sb(\omega _s)e^{-i\omega _s(t-t_{0k})}},
\end{eqnarray}
where $\eta _{2(3)}$ is quantum efficiency of SPD2 (SPD3).

Because the individual signal and idler fields generated by the SFWM parametric process are thermal states, the joint state of signal and idler fields, originated from the $k$th and $l$th pump pulse, is a zero-mean Gaussian state. Therefore, the second order moments of the joint state are given by its temporal auto- and cross-correlation functions as follows:
\begin{eqnarray}
\label{15}
  \left\langle 0 \right|E_{j}^- (t_1-t_{0k})E_{j}^ +  (t_2-t_{0l} )\left| 0 \right\rangle & \propto &  \left| G \right|^2 \int {d\omega _{s(i)} d\omega _{s(i)} 'f_{s(i)}^  *  (\omega _{s(i)} )f_{s(i)}(\omega _{s(i)} ')e^{i\omega _{s(i)} (t_1-t_{0k} ) - i\omega _{s(i)} '(t_2-t_{0l})  }} \cr
   &\times &  \int {d\omega_{i(s)} \phi ^* (\omega_{i(s)} ,\omega _{s(i)} )}  \phi (\omega_{i(s)} ,\omega _{s(i)} ') ~~~~(j=1,2,3),
\end{eqnarray}
\begin{eqnarray}\label{16}
\left\langle 0 \right|E_1^ +  (t_1-t_{0k} )E_{2(3)}^ +  (t_2-t_{0l} )\left| 0 \right\rangle  \propto G\int {d\omega _s d\omega _i f_s (\omega _s )f_i (\omega _i )e^{ - i\omega _s (t_1-t_{0k})  - i\omega _i (t_2-t_{0l}) } \phi (\omega _s ,\omega _i )},
\end{eqnarray}
and
\begin{eqnarray}\label{17}
   \left\langle 0 \right|E_1^-  (t_1-t_{0k})E_{2(3)}^ +  (t_2-t_{0l})\left| 0 \right\rangle  = \left\langle 0 \right|E_j^ +  (t_1-t_{0k})E_{j}^ +  (t_2-t_{0l} )\left| 0 \right\rangle  = 0.
\end{eqnarray}

\subsection{Photon statistics}

In the process of analyzing the second-order intensity correlation of the heralded single photons, we first find out the count probability of SPD1, SPD2, and SPD3, respectively; then, to determine the reliability of HSPS, we calculate the two-fold coincidence between SPD1 and SPD2 (or SPD3) to quantify the quantum correlation of photon pairs; finally, to show the capability of creating one and only one photon per heralding events, we compute
intensity correlation function for the signal field, conditioned on observing an idler photon count.

According to Eqs (\ref{13})-(\ref{14}), the
count probability per pulse of SPD1 and SPD2 (SPD3) can be written as:
\begin{eqnarray}\label{18}
P_{1}=\int_{{- T_r /2}}^{{T_r /2}}{dt\left\langle 0\right\vert
E_{1}^-(t-t_{0k})E_{1}^{+}(t-t_{0k})\left\vert 0\right\rangle }=\sqrt{2}\pi \left\vert G\right\vert ^{2}\eta _{i}\eta _{1}\frac{{\sigma _{i}}}{{\sigma _{p}}}
\end{eqnarray}
and
\begin{eqnarray}\label{19}
P_{2(3)}=\int_{{- T_r /2}}^{{ T_r /2}}{dt\left\langle 0\right\vert
E_{2(3)}^-(t-t_{0k})E_{2(3)}^{+}(t-t_{0k})\left\vert 0\right\rangle }=\frac{{\pi}}{{\sqrt{2}}} \left\vert G\right\vert ^{2}\eta _{s}\eta _{2(3)}\frac{{\sigma _{s}}}{{\sigma _{p}}},
\end{eqnarray}
where $T_{r}$ is the response time of SPD, determined by its corresponding gate width.

We note that the count probability of any SPDs is a time integral from $- T_r /2 $ to $T_r /2 $. Because the gate width of SPDs is much longer than the creation time period of photon pairs confined within the pulse duration, $P_{1}$ and $P_{2(3)}$ in Eqs. (\ref{18}) and (\ref{19}) can be treated as a integral from $-\infty $ to $\infty$.

The two-fold coincidence count probability per pulse between SPD1 and SPD2 (SPD3) can be expressed as
\begin{eqnarray}\label{20}
P_{12(3)}(t_{0k},t_{0l})=\int_{{- T_r /2}}^{{+ T_r /2}}{\int_{{- T_r /2}}^{{+ T_r /2}}} {dt_{s}dt_{i}\left\langle 0\right\vert
E_{2(3)}^-(t_{s}-t_{0k})E_{1}^-(t_{i}-t_{0l})E_{1}^{+}(t_{i}-t_{0l})E_{2(3)}^{+}(t_{s}-t_{0k})\left\vert
0\right\rangle }.
\end{eqnarray}
We carry out the calculation by using the quantum form of the Gaussian moment-factoring
theorem. For the signal and idler photons generated from the different time slots of pump pulses, we have $t_{0k}-t_{0l}\neq0$ and $t_{0k}-t_{0l}\gg T_r$. In this case, the coincidence count probability is
\begin{eqnarray}\label{21}
P_{12(3)}(t_{0k},t_{0l})=P_{1}P_{2},
\end{eqnarray}
which is originated from the multi-photon events, and is referred as the accidental coincidence.
While for
 the signal and idler photons produced in the same time slot, i.e., $t_{0k}=t_{0l}$, we arrive at
\begin{eqnarray}\label{22}
P_{12(3)}(0)=P_{1}P_{2}+\frac{{1}}{{2}}\eta _{s}\eta _{2(3)}P_{1}\xi _{s},
\end{eqnarray}
where $\xi _{s}=\frac{{\sigma _{s}}}{\sqrt{2\sigma
_{p}^{2}+\sigma _{i}^{2}+\sigma _{s}^{2}}}$ is the collection efficiency of the photon pairs for the heralding photons detected in idler channel~\cite{li10}, and  $0$ in the bracket stands for $t_{0k}-t_{0l}=0$. Eq. (\ref{22}) indicates the coincidence of signal and idler photons produced by the same pump pulse is the sum of the accidental coincidence and the true coincidences come from the pair events.

In carrying out the calculation in Eq. (\ref{22}), the result of the time integral in Eq. (\ref{20}) becomes a $\delta $-function. Moreover, considering the quantum correlation only exist between the signal and idler photons produced in the same pulse, for the correlation hereinafter calculated and verified, we let $t_{0k}=t_{0l}=0$, and treat the time integral from $- T_r /2 $ to $T_r /2 $ as that from $-\infty $ to $\infty$.

With the expressions of the two-fold coincidence count probability, we can analyze the quantum correlation of the photon pairs, $g_{si}^{(2)}(t_1,t_{2})$, which is determined by the degree of second-order intensity correlation between the signal and the idler fields
\begin{eqnarray}\label{23}
g_{si}^{(2)}(t_1,t_{2})=\frac{\langle {E_{2(3)}^-(t_{2})E_1^-(t_1)E_1^{+}(t_1)E_{2(3)}^{+}(t_{2})}\rangle
}{\langle {E_1^-(t_1 ) E_1^{+}(t_1 )\rangle \langle E_{2(3)}^-(t_{2})E_{2(3)}^{+}(t_{2} )}\rangle }.
\end{eqnarray}
Eq. (\ref{23}) shows that the precise knowledge of $g_{si}^{(2)}(t_1,t_{2})$ needs SPDs with a resolution time shorter than the duration of pump pulse. However, in practice, limited by the resolution time $T_r$, the measurement of $g_{si}^{(2)}(t_1,t_{2})$ is a time integral from $- T_r /2 $ to $T_r /2 $, i.e. $\left\langle {g_{si}^{(2)} (t_1, t_2 )} \right\rangle _{T_r } = \frac{{P_{12(3)}(0)}}{{P_{1}P_{2(3)}}}$. Since the coincidence to accidental-coincidence ratio $CAR= \frac{{P_{12(3)}(0)}}{{P_{1}P_{2(3)}}}$ is a directly measurable quantity, we will use $CAR$ to quantify the temporal correlation of photon pairs in the following description.

Defining the normalized detection rate of photon pairs as  $P_{pair}=P_{1}\xi _{s}/(\eta _{i}\eta _{1})$, and using Eqs (\ref{18}), (\ref{19}) and (\ref{22}), we obtain the expression of the coincidence to accidental-coincidence ratio:
\begin{eqnarray}\label{24}
CAR=\left\langle {g_{si}^{(2)} (t_1, t_2 )} \right\rangle _{T_r }=1+\frac{{\sigma _{s}^{\prime }\sigma _{i}^{\prime }}}{{P_{pair}(2+\sigma
_{s}^{\prime 2}+\sigma _{i}^{\prime 2})}},
\end{eqnarray}
where $\sigma _{s(i)}^{\prime }=\sigma _{s(i)}/\sigma _{p}$. It is clear that $CAR$ decreases with the increase of $P_{pair}$, which determines the brightness of HSPS. To further demonstrate the dependence of $CAR$, we then plot $CAR$ as a function of the bandwidth of signal and idler photons. As shown in Fig. 2, at a certain rate of $P_{pair}$, $CAR$ obviously increases with the increase of  $\sigma _{s(i)}^{\prime }$. Looking at Fig. 2 more carefully, we find that for the idler (signal) photons shaped by $\sigma _{i(s)}^{\prime }$, if the bandwidth of signal (idler) photon $\sigma _{s(i)}^{\prime }$ is less than $\sqrt{ 2+ \sigma _{i(s)}^{\prime 2}}$, $CAR$ increases with the increase of $\sigma _{s(i)}^{\prime }$; otherwise, $CAR$ decreases with the increase of $\sigma _{s(i)}^{\prime }$. Therefore, the relation described by Eq.(\ref{24}) is the basis for optimizing the quantum correlation between signal and idler photon pairs, which is the key to reliably generate heralded single photons.

To obtain the second order intensity correlation function of the HSPS, it is necessary to know the theoretical expression of the triple coincidence probability between SPD1, SPD2 and SPD3 in the same time slot:
\begin{eqnarray}\label{25}
P_{123}(0)&=&\int {dt_{1}dt_{2}dt_{3}\left\langle 0\right\vert
E_{1}^-(t_{1})E_{2}^-(t_{2})E_{3}^-(t_{3})E_{3}^{+}(t_{3})E_{2}^{+}(t_{2})E_{1}^{+}(t_{1})\left\vert 0\right\rangle
}\cr
&=&\int {dt_{1}\left\langle 0\right\vert
E_{1}^-(t_{1})E_{1}^{+}(t_{1})\left\vert 0\right\rangle }\int {%
dt_{2}dt_{3}\left\langle 0\right\vert
E_{2}^-(t_{2})E_{3}^-(t_{3})E_{3}^{+}(t_{3})E_{2}^{+}(t_{2})\left\vert
0\right\rangle } \cr
&+&2\int {dt_{2}\left\langle 0\right\vert
E_{2}^-(t_{2})E_{2}^{+}(t_{2})\left\vert 0\right\rangle }\int {dt_{1}dt_{3}}%
\left\vert {\left\langle 0\right\vert E_{1}^-(t_{1})E}_{{3}}^-{(t_{3})\left\vert
0\right\rangle }\right\vert ^{2} \cr
&+&2\int {dt_{1}dt_{2}dt_{3}\left\langle 0\right\vert
E_{1}^-(t_{1})E_{2}^-(t_{2})\left\vert 0\right\rangle \left\langle 0\right\vert
E_{3}^-(t_{3})E_{2}^{+}(t_{2})\left\vert 0\right\rangle \left\langle
0\right\vert E_{3}^{+}(t_{3})E_{1}^{+}(t_{1})\left\vert 0\right\rangle } \cr
&=&P_{1}P_{2}P_{3}+\eta _{s}\xi _{s}P_{1}\frac{{\eta _{3}P_{2}+\eta _{2}P_{3}}}{{2}}+( {g_s^{(2)}-1})(P_{1} P_{2}P_{3}+ \eta _{s}\xi _{s}^{\prime }P_{1}\frac{{\eta _{3}P_{2}+\eta _{2}P_{3}}}{{2}}),
\end{eqnarray}
where $g_s^{(2)}=1+\frac{1}{\sqrt{1+\frac{%
\sigma _{s}^{\prime 2}}{2}}}$ can be considered as the normalized second order intensity correlation function of individual signal field~\cite{li08ol}, and the coefficient $\xi _{s}^{\prime }=\frac{\sqrt{2}\sigma _{s}^{\prime }}{{\sqrt{4+2\sigma _{i}^{\prime 2} +\sigma _{s}^{\prime 2}}}}$ means the collection efficiency to the two pairs events. One sees that triple coincidence probability $P_{123}(0)$ is a sum of the three terms: the first term is the accidental coincidence between SPD1, SPD2 and SPD3; the second term stems from the coincidence between a signal photon and a photon pairs; and the third term is associated with the bunching effect in the heralded signal band. If the bandwidth of the heralded signal $\sigma _{s}^{\prime }$ is very broad, the sum of the first two terms would be a good proximation of $P_{123}(0)$; if not, however, the bunching effect must be taken into account.

Having obtained the expression of the various counting probabilities of the SPDs, we are able to compute the second order
intensity correlation for the heralded signal field, conditioned on observing an idler photon count in the time slot $t_{0k}$, i.e.,
\begin{eqnarray}\label{26}
g_c^{(2)}(0)=\frac{P_{123}(0)P_{1}}{P_{13}(0)P_{12}(0)}.
\end{eqnarray}
Substituting Eqs. (\ref{18}), (\ref{22}) and (\ref{25}) into Eq.(\ref{26}), and using the approximation $\frac{1}{{g_s^{(2)}}}+\frac{{g_s^{(2)}-1}}{{g_s^{(2)}}}\frac{{\xi_{s}^{\prime }}}{{\xi _{s}}}\approx 1$~\cite{note01}, we have
\begin{eqnarray}\label{27}
g_c^{(2)}(0) \approx \frac{{g_s^{(2)}
}}{{CAR}}\left( {2-\frac{1}{{CAR}}}\right).
\end{eqnarray}
Eq. (\ref{27}) shows that the value of $g_c^{(2)}(0)$ is determined by the second order correlation function of the individual heralded signal field and quantum temporal correlation of the photon pairs.

Although the condition $g_c^{(2)}(0)\approx 0$ is desirable for a single photon source, the desirability
can not be satisfied unless $CAR\gg 1$. To achieve $CAR\gg 1$, according to Eq. (\ref{24}), one should either
decrease the brightness characterized by $P_{pair}$, or increase $\sigma _{s}^{\prime }$ and $\sigma _{i}^{\prime }$.
Because a decreased brightness is adverse for the applications
of HSPS, to figure out how to optimize the photon statistics, we plot $g_c^{(2)}(0)$ as a function of the
bandwidth of signal and idler photons for $P_{pair}=0.02$ pairs/pulse.
As shown in Fig. 3, the value of
$g_c^{(2)}(0)$ decrease with the increase of $\sigma _{s}^{\prime }$ and $\sigma _{i}^{\prime }$, and
the asymmetry, owing to the bunching effect of heralded signal photons, can be observed in the pattern of the contour-plot.

\subsection{Heralding efficiency}

In addition to the intensity correlation for the heralded signal field, another important parameter to characterize a HSPS is the heralding efficiency, $H$. Given the detection of one member of a photon pair, $H$ is the probability that the twin photon is actually
present in the output of HNLF, which can be obtained by
measuring the conditional detection efficiency of
heralded photons, and then correcting for the
losses in the heralded-photon
analysis zone (the shaded area shown in Fig. 1). If the conditional
detection efficiency $\eta_D$  is simply defined
as the ratio between true coincidence and
the trigger photon detection rate, i.e.
\begin{eqnarray}\label{28}
\eta _D=\frac{P_{12(3)}(0)-P_1P_{2(3)}}{P_1}=\frac 12\eta _s\eta _{2(3)}\xi _s,
\end{eqnarray}
the heralding efficiency can be expressed as
\begin{eqnarray}\label{29}
H=\xi _s=\frac{{\sigma _{s}}^{\prime}}{\sqrt{2+{\sigma _{i}^{\prime}}^{2}+{\sigma _{s}^{\prime}}^{2}}}.
\end{eqnarray}
Eq. (\ref{29}) shows the heralding efficiency $H$ is equivalent to the collection efficiency of the photon pairs for the heralding trigger photons detected in idler channel~\cite{li10}. Figure 4 plots $H$ as a function of the bandwidth of signal and idler photons. One sees that the broader the bandwidth of F2 in signal field relative to that of the filters in pump and idler fields, the higher the heralding efficiency.

\subsection{Indistinguishability}

Apart from the photon statistics and heralding efficiency, indistinguishability of HSPS is important as well. Because of the independent nature
of the photons, in order to obtain high visibility
in quantum interference between multiple sources, photons must be in identical single mode so
that they are indistinguishable. Although
the spontaneous parametric emission processes are independent in different sources,
the timing provided by the ultrashort pump
pulse can be used to realize the required synchronization. For the photon pairs generated in fiber based sources, spatial overlap is ensured, spectral overlap can be
achieved by the use of similar filters, and polarization can be matched by a polarization controller. However, the spectral of signal and idler photon are usually correlated, and the temporal mode structure is messed up~\cite{li08ol,li08a-opex}.
To preserve the temporal indistinguishability,
optical filtering is necessary so that the detected signal and idler photons are spectrally factorable.

According to the field operators of signal and idler photons shown in Eqs. (\ref{11}) and (\ref{12}), photon pairs in spectral factorable state can be obtained by using two methods. One method is applying a narrow band filter (NBF) in idler band, so that the heralding photons can be viewed as in a single mode. For example, when bandwidth of the filter in heralding idler band is $\sigma _{i}^{\prime }=0.3$, the second order coherence function of the individual field $g_i^{(2)}=(1+\frac{1}{\sqrt{1+\frac{%
\sigma _{i}^{\prime 2}}{2}}})$ is about 1.98~\cite{li08ol}, indicating the temporal mode of the individual idler photons can be considered as a single mode~\cite{Ou99}. Under this condition, no matter what kind of filter is applied in signal band, the HSPS is indistinguishable. The other method is inserting a NBF in signal band, so that the heralded photons is in a single mode. For example, when the filter in heralded signal band is $\sigma _{s}^{\prime }=0.3$, the HSPS is indistinguishable, with no relevant to the filter in idler band. Substituting the source parameters given by the two examples into Eqs. (\ref{24}), (\ref{27}) and (\ref{29}), we calculate the corresponding $g_c^{(2)}(0)$ and $H$ for the pair rate $P_{pair}=0.005$ pairs/pulse. As shown in Fig. (\ref{indistinguishability}), in the sense of evaluating the source by looking at the value of $g_c^{(2)}(0)$ and $H$, indistinguishable HSPS obtained by using NBF
in the heralding band is slighter better, because in this case, the contribution of the bunching effect of the heralded signal photons (see Eq. (25)) can be made smaller, and heralding efficiency can be adjusted to be higher by broadening the bandwidth of heralded signal photons.

It is well known that indistinguishable HSPS realized by using NBFs will severely limit the brightness of single photons. However, we would mention that the even worse characteristic of the HSPS is the increased value of $g_c^{(2)}(0)$. Comparing Fig. 5 with Fig. 3, we find that although the brightness in Fig. 5 is only $1/4$ of that in Fig. 3, the smallest $g_c^{(2)}(0)$ achievable in Fig. 5 is higher than that in Fig. 3. This is because the bandwidth setting of signal and idler photon for a indistinguishable HSPS are not optimized for obtaining a maximized $CAR$.
For the individual signal and idler photons, the portion that their twin photons are not within the bandwidth
of idler (signal) filter is not minimized. Moreover, the indistinguishable HSPS achieved by using a NBF in heralded band, will additionally suffer from the low heralding efficiency. Therefore, to develop a indistinguishable heralded single photons with high quality, direct generation of photon pairs in spectral factorable state is desirable~\cite{Garay-Palmett-OE2007}.

Because we have previously verified the indistinguishability of a fiber based HSPS by using NBF in heralded band and experimentally studying the forth-order interference between independent sources, as demonstrated in Ref.~\cite{li08ol} and \cite{li08b-opex}, we will not discuss this issue in the experiments described in the next section.

\subsection{Interplay of photon statistics, heralding efficiency and indistinguishability}

From the above analysis of the dependence of photon statistics, heralding efficiency and indistinguishability, one sees the three criteria of the HSPS can not be optimized simultaneously. For a HSPS with a certain brightness, i. e., the pair rate $P_{pair}$ is kept at a certain level, according to Eqs. (\ref{27}) and (\ref{29}), enlarging the bandwidth of signal and idler photons helps to reduced $g_c^{(2)}(0)$ and increase $H$, but with a cost of decreased indistinguishability. On the other hand, to obtain a good indistinguishability, a NBF in signal or idler band is required. This will result in a relatively increased  $g_c^{(2)}(0)$, although $H$ with a higher value can be maintained by optimizing the source parameters.

It is worth noting that the brightness of HSPS, characterized by $P_{pair}$, is also one of the concerns for exploring its practical applications. There are two approaches to make the source brighter: one is using signal and idler photons with a broader bandwidth (see Eqs, (\ref{18})-(\ref{22})), the other is to increase the gain of SFWM by increasing the pump power and the length of fiber. The former will result in a HSPS with a relatively low  $g_c^{(2)}(0)$ and higher $H$. However, in addition to a poor indistinguishability, this kind of source is not suitable for studying the Fock state with the photon number greater than 1, since the gain of SFWM (per mode) is low, and the two-pairs events is relatively rare. While the latter will lead to a increased $g_c^{(2)}(0)$. But this kind of source is useful for the investigations involving the multi-photon pairs events. Moreover, to describe the latter more precisely, the field operators of signal and idler photons shown in Eq. (11) should take a higher order approximation.

\section{Experiments and results}

A schematic of our experimental setup is shown in
Fig. 6. Signal and idler photon pairs at wavelengths of
1544.53 nm  and 1531.9 nm, respectively, are produced by launching a pulsed pump centering at 1538.9 nm along the fast or slow axis of the 20-meter-long HNLF, whose zero-dispersion wavelength, dispersion slope, and birefringence ($\Delta n$) are measured to be about $1525 \pm 5$ nm, 0.02 ps/nm$^2\cdot$km, and $2.5\times 10^{-6}$, respectively. The fiber polarization controller (FPC1) is used to adjust the polarization of pump. The HNLF with $\gamma=11 W^{-1}km^{-1}$ is immerged in liquid nitrogen to suppress Raman scattering. FPC2 and fiber polarization beam splitter (FPBS) placed at the output port of HNLF are used to select the co-polarized photon pairs and to reject the residual Raman scattering which is cross-polarized. The loss of HNLF mainly comes from fusing loss (~0.8dB) due to core diameter mismatch between HNLF fiber and spliced standard single-mode fiber.

The pump with a 10-ps-duration and a repetition rate of about 41 MHz is spectrally carved out from a mode-locked femto-second fiber laser. To achieve the required power, the pump pulses are amplified by an erbium-doped-fiber amplifier (EDFA), photons at the signal and
idler wavelength from the fiber laser that leak through the spectral-dispersion optics and from the amplified spontaneous emission from the EDFA are suppressed by passing the
pump through a tunable filter F1 with a FWHM of about 0.3 nm.

The dual-band filter F2, providing a pump-rejection ratio in excess of 120 dB, is obtained by cascading four WDM filters (Santec, model WDM-15). To investigate the influence of the bandwidth of detected signal and idler photons, F2 is realized by using different combination of 100 GHz and 200 GHz WDM filters, so that the FWHM in signal and idler bands could be either 0.6 and 0.6 nm, or 1.1 and 0.6 nm, or 1.1 and 1.1 nm, respectively. For convenience, we label the three kinds of F2 as $F2_I$, $F2_{II}$ and $F2_{III}$, whose spectra are plotted in inset of Fig. 6.

The quantum efficiencies of SPD1, SPD2 and SPD3 are about $20\%$, $12\%$ and $17\%$, respectively, whose corresponding dark-count probabilities are about $1.9*10^{-5}$, $2.1*10^{-5}$ and $5.9*10^{-5}$ counts/pulse, respectively.
The electrical signals produced by the SPDs  in response to the incoming photons (and dark counts) are
acquired by a photon-counting system. Thus, single counts in both
the signal and idler bands and coincidences acquired from different time slots can be
recorded. To determine the total detection efficiencies for idler and signal channels, besides the efficiency of the corresponding SPDs, the efficiencies of the HNLF $70\%$, the filter $F2$ (including $F2_I$, $F2_{II}$ and $F2_{III}$) in signal and idler channels, $\sim 24\%$ and $\sim 52\%$, 50/50 fiber coupler, and other transmission components about $\sim 90\%$ should be taken into account.

Before characterizing the HSPS, we first describe the quantum correlation of photon pairs generated in HNLF by using SPD1 and SPD2 to record single counts and coincidence rate when dual band filter is set to be $F2_I$, $F2_{II}$ and $F2_{III}$, respectively. Figure 7(a) shows the number of scattered
photons in idler band per pump pulse, $N_{i}$ as a function of the
average pump power, $P_{ave}$, and
the measured data is fitted with the $N_{i}=s_1P_{ave}+s_2P_{ave}^2$,
where $s_1$ and $s_2$ are the linear and quadratic coefficients,
which respectively determine the strengths of RS and FWM in HNLF. Figure 7(b) shows the measured coincidence and accidental coincidence
rates of signal and idler photons, produced by same pump pulse and adjacent pump pulses, respectively, as a function of the average pump power. It is clear, the quadratic part $s_2P_{ave}^2$ in Fig. 7(a) is equivalent to $P_1$ in Eq. (18). Thus, for each kind of dual band filter, we can deduce the rate of signal and idler photons via SFWM, the normalized detection rate of photon pairs, $P_{pair}$, and the ratio between the coincidence and accidental coincidence rates, $CAR$. Figure 7(c) shows the value of $CAR$ versus $P_{pair}$ after the effect of Raman scattering is subtract. One sees that for a certain value of $P_{pair}$, $CAR$ increases with the increase of the bandwidth in the detected signal and idler fields. This is consistent with the theory predictions in Fig. 2.

We then use the detection signal of SPD1 to herald the existence of single photons in signal band, and measure its conditional second order correlation function $g_c^{(2)}(0)$ by recording the single counts and coincidences between SPD1, SPD2, and SPD3 at various conditions. When the detected photons of the three SPDs originate from the same pump pulse, we record the triple coincidence at different pump power levels by using $F2_I$, $F2_{II}$ and $F2_{III}$, respectively. During the measurement, the single counts of SPD1, SPD2, and SPD3,
and two-fold coincidences between two arbitrary SPDs are also recorded. The corresponding value of $g_c^{(2)}(0)$ can be obtained according to Eq. (\ref{26}). To figure out the dependence of $g_c^{(2)}(0)$ upon the brightness in different cases, we extract the normalized detection rate of photon pairs, $P_{pair}$ from the measured single counts and two-fold coincidences, and sketch $g_c^{(2)}(0)$ as a function of $P_{pair}$. As shown in Fig. 8 (a), for a certain $P_{pair}$, $g_c^{(2)}(0)$ obtained by using $F2_{III}$, with FHWM of 1.1 nm in both signal and idler channels tends to be the smallest, while $g_c^{(2)}(0)$ obtained by using $F2_{I}$, with FHWM of 0.6 nm in both signal and idler channels, tends to be the biggest. The result is qualitatively in accordance with the theory expectation in Fig. 3. We note that with the reduction of $P_{pair}$, the discrepancy between the three sets of data become unapparent. This is because the contamination of Raman scattering is getting more severe with the reduction of the pump power and with the broadening of the bandwidth of signal and idler photons, as illustrated by the fitting parameters and curves in Fig. 7(a). Therefore, correcting the measured results of $g_c^{(2)}(0)$ by subtracting the influence of Raman scattering, we plot the $g_c^{(2)}(0)$ as a function of $P_{pair}$, and compare the modified data with the calculated results. The calculation is carried out by substituting the source parameters into Eqs. (\ref{24}) and (\ref{27}), in which no fitting parameter is used, and the FWHMs of the Gaussian shaped filters are chosen to be the same as that of the Super Gaussian shaped F2. As shown in Fig. 8 (b), the corrected $g_c^{(2)}(0)$ agrees with calculated results, although there is departure between them, which is originate from the differentia of the filters used in experiments and theoretical model, respectively.

Finally, we characterize the heralding efficiency of the HNLF-based HSPS. Using the data in Fig. 7(b), we deduce the true coincidence, $R_{true}$, which is the difference of coincidence and accidental coincidence rate between SPD1 and SPD2. To obtain the heralding efficiency $H$, according to Eqs. (\ref{28})-(\ref{29}), we first divide $R_{true}$ by the single counts of SPD1, which is the trigger photon detection rate, and then correct the result by using the total detection efficiency in the signal channel connected with SPD2. Figure 9(a) plots the measured $H$ as a function of pump power for signal and idler photons with different bandwidth combinations. One sees at a certain power, the biggest and smallest value of $H$ can be achieved by using signal and idler photons shaped by $F2_{II}$ and $F2_{I}$, respectively. In the sense of comparing the relative value of $H$ obtained in different cases, the results qualitatively agrees with Eq. (\ref{29}). However, instead of being irrelevant to pump power, it is obvious that for signal and idler photons with a certain spectra, $H$ increases with the increase of pump power. The departure is caused by the contamination of Raman scattering. To confirm this point, using the fitting results in Fig. 7(a), we modified the results by subtracting the influence of Raman scattering and compare them the calculated results obtained by substituting the source parameters into Eq. (\ref{29}). As shown in Fig. 9(b), for the signal and idler photons with a specified spectra, the modified $H$ does not depend on the pump power, and agree with theory expectations.

\section{Conclusion}

In conclusion, using the photon pairs produced in 20-meter long HNLF by a pulsed pump in 1550 nm band, we have studied the HSPS from the theoretical and experimental aspects. Based on the Bogoliubov transformation for a broadband pulsed pump, we derive the analytic expressions of the second-order coherence function $g_c^{(2)}(0)$ and heralding efficiency $H$ of the HSPS for the first time. The calculation reveals the dependence of $g_c^{(2)}(0)$, $H$ and indistinguishability upon the source parameter, and the validity of the theoretical calculations are proved by the experimental results. We believe the theoretical frame used to model the fiber based HSPS is also useful for describing other spontaneous parametric process pumped with a mode-locked laser.

The HSPS presented in this paper is an all-fiber source with compact size and freedom from misalignment. Although heralded single photons in 1550 nm telecom band are desirable for quantum communication, we would like to mention that limited by the detection techniques, heralding photons in visible band are more practical at the current stage~\cite{mcmillan09}. However, the use of 1550 nm heralding photons brings us a lot of conveniences in investigating the dependence of $g_c^{(2)}(0)$ and $H$ under different experimental conditions, because various kinds of off-the-shelf fiber components with high quality and low cost are commercially available in this band.

We think the HSPS can be improved from the following aspects. Firstly, the discrepancies between the directly measured experimental results and theory calculations due to the existence of Raman scattering can be reduced or wiped off by cooling the HNLF to 4 K~\cite{Nam08OPEX} or by using photonic crystal fiber based photon pairs with large detuning to further mitigate Raman effect~\cite{Fan08,cohen09,mcmillan09,Alibart06}. Secondly,
the current transmission loss experienced by heralded signal photons can be reduced to less than 1 dB by minimizing splicing loss and using high quality WDM or fiber-bragg-grating filters. Thirdly, the emission rate of the heralded single photons can be
be dramatically increased by pumping the HNLF with a mode-locked
fiber laser operating at a repetition rate above
10 GHz and by using SPDs with high speed and high efficiency~\cite{Korneev04,Namekata09}. Finally,
to further increase the single photon rate while maintaining a low value $g_c^{(2)}(0)$, similar to multiplexed SPDC scheme used to better approximate a single photon source on demand~\cite{migdall02},
it is possible to operate an array of simultaneously pumped SFWM sources. In this case, in the sense of obtaining a reduced $g_c^{(2)}(0)$ and increased $H$ at a certain pair rate $P_{pair}$ for each SFWM source, our study on characterizing the HNLF based HSPS would be useful for optimizing parameters.

\begin{acknowledgments}
We would like to thank Prof. Z. Y. Ou for useful discussion. This work was supported in part by the NSF of China
(No. 10774111), Foundation for Key Project of
Ministry of Education of China (No. 107027), the Specialized Research Fund for the Doctoral Program of Higher Education of China (No. 20070056084), 111 Project B07014, and the State Key
Development Program for Basic Research of China (No. 2010CB923101)

\end{acknowledgments}


\begin{thebibliography}{32}
\bibitem[{not()}]{note01}
\bibinfo{journal}{The approximation can not be further simplified. If the
  filter applied in signal field is narrow band, $\frac{{\xi_{s}^{\prime
  }}}{{\xi _{s}}}=1 $; if the applied filter is broadband,
  $\frac{{\xi_{s}^{\prime }}}{{\xi _{s}}}\neq 1$, but the value of $g_s^{(2)}$
  become smaller, the approximation still holds}.
\end{thebibliography}

\newpage

\begin{figure}[htbp]
\centering
\includegraphics[width=8cm]{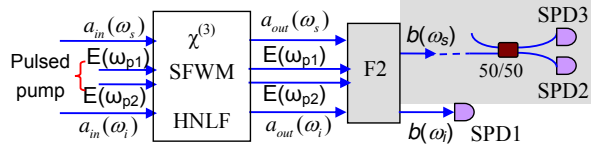}
\caption{(Color online) Conceptual representation of the scheme for generating heralded single photons. F2, dual-band filter.}
\label{setup}
\end{figure}

\begin{figure}[htbp]
 \centering
\includegraphics[width=4.5cm]{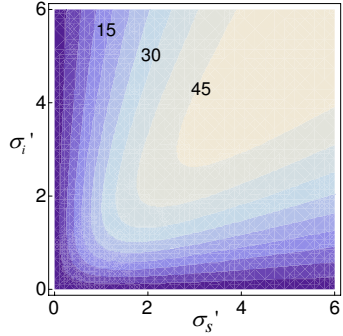}
  \caption{(Color online) Contour-plot of quantum correlation of photon pairs $CAR$ as a function of the bandwidth of signal and idler photons. The calculation is carried out at the pair rate $P_{pair}=0.01$ pairs/pulse, and $\sigma _{s(i)}^{\prime }=\sigma _{s(i)}/\sigma _{p}$ }
\end{figure}

\begin{figure}[htbp]
  \centering
  \includegraphics[width=5cm]{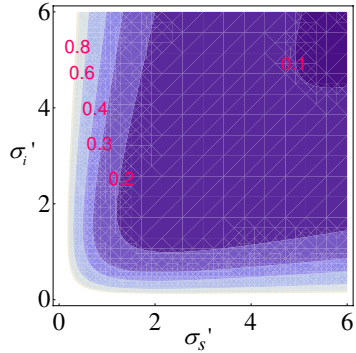}
  \caption{(Color online) Contour-plot of $g_c^{(2)}(0)$ as a function of the
bandwidth of signal and idler photons for $P_{pair}=0.02$ pairs/pulse. }
\end{figure}

\begin{figure}[htbp]
  \centering
  \includegraphics[width=4.5cm]{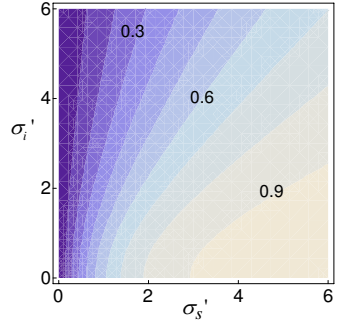}
  \caption{(Color online) Contour-plot of heralding efficiency of the heralded single photons in signal field $H$ as a function of the bandwidth of signal and idler photons. }
\end{figure}

\begin{figure}[htbp]
\centering
 \includegraphics[width=8cm]{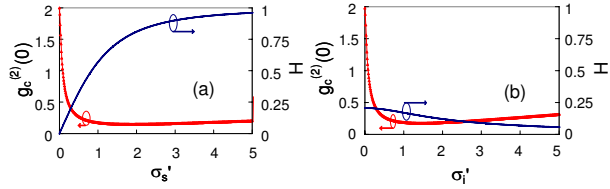}
\caption{(Color online) $g_c^{(2)}(0)$ and $H$ of indistinguishable heralded single photons with a pair rate of  $P_{pair}=0.005$ pairs/pulse for (a) heralding photons is in a single mode, realized by setting $\sigma _{i}^{\prime }=0.3$, and (b) heralded photons is in a single mode, achieved by setting $\sigma _{s}^{\prime }=0.3$.  }
 \label{indistinguishability}
\end{figure}

\begin{figure}[htbp]
  \centering
\includegraphics[width=8cm]{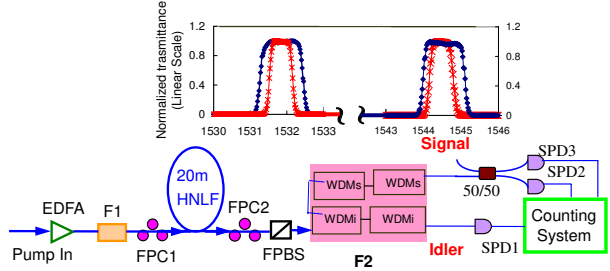}
  \caption{(Color online) Schematic of the experimental setup. The inset is the spectra of dual band filter F2, stars and diamonds represent the spectra with FWHM of 0.6 and 1.1 nm, respectively.   }
\end{figure}

\begin{figure}[htbp]
\centering
\includegraphics[width=8cm]{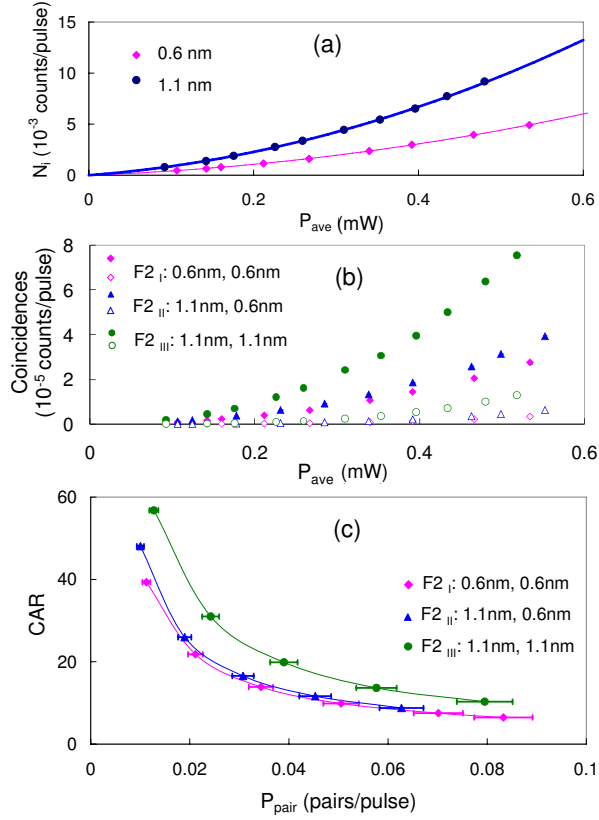}
\caption{(Color online) (a) The number of scattered photons in idler band per pulse, $N_{i}$, as a function of the average pump power, $P_{ave}$. Thick and thin solid curves are the fitting of $N_{i}=s_1P_{ave}+s_2P_{ave}^2$, with $s_1=0.061$, $s_2=0.027$, and $s_1=0.030$, $s_2=0.012$, for F2 in idler band with FWHM of 1.1 and 0.6 nm, respectively. (b) Coincidences rates (represented by solid circles, triangles and diamonds) and accidental coincidence rate (represented by hollow circles, triangles and diamonds) of the detected signal and idler photons versus $P_{ave}$. (c) The deduced ratio between the coincidence and accidental coincidence rates, $CAR$, as a function of the normalized photon pair detection rate, $P_{pair}$, after the influence of Raman scattering is excluded. The solid curves are only for guiding eyes.}
\end{figure}

\begin{figure}[htbp]
  \centering
  \includegraphics[width=8cm]{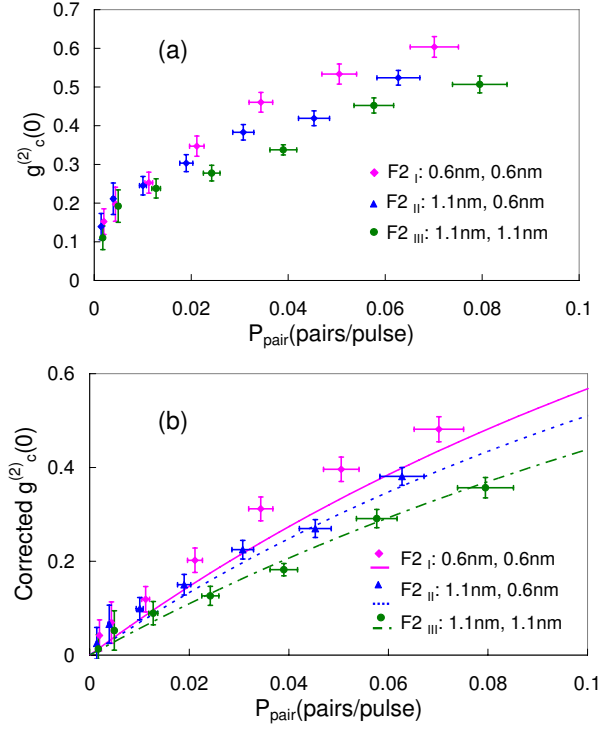}
  \caption{(Color online) (a) Measured and (b) corrected $g_c^{(2)}(0)$ of the HNLF-based HSPS versus brightness, characterized by the normalized photon pair detection rate $P_{pair}$. The size of the error bar of $P_{pair}$ is determined by the uncertainty of the total detection efficiencies, and the uncertainty of $g_c^{(2)}(0)$ lies on the statistical fluctuation of photon counting. The theory curves in (b) are calculated by using Gassian shaped filters with FWHM the same as that of F2.}
\end{figure}

\begin{figure}[htbp]
  \centering
  \includegraphics[width=8cm]{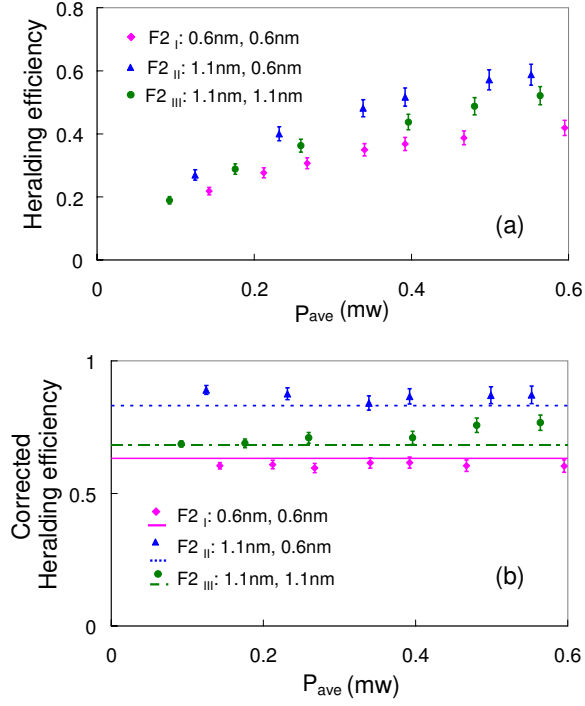}
  \caption{(Color online) (a)Heralding efficiency of HNLF-based HSPS, $H$, as a function of average pump power. (b) The corrected heralding efficiency versus the average pump power when the Raman photons in idler bands are subtracted. The straight lines in (b) are the theoretically calculated results, in which the FWHMs of Gaussian shaped filters are the same as that of F2.}
\end{figure}

\end{document}